# Facile fabrication of suspended as-grown carbon nanotube devices

V. K Sangwan[1,2], V. W. Ballarotto[2], M. S. Fuhrer[1], E. D. Williams[1,2]

1. Department of Physics, University of Maryland, College Park, MD 20742

2. Laboratory for Physical Science, College Park, MD 20740

**Abstract:**

A simple scalable scheme is reported for fabricating suspended carbon nanotube field effect transistors (CNT-FETs) without exposing pristine as-grown carbon nanotubes to subsequent chemical processing. Versatility and ease of the technique is demonstrated by controlling the density of suspended nanotubes and reproducing devices multiple times on the same electrode set. Suspending the carbon nanotubes results in ambipolar transport behavior with negligible hysteresis. The Hooge's constant of the suspended CNT-FETs ($2.6 \times 10^{-3}$) is about 20 times lower than for control CNT-FETs on $SiO_2$ ($5.6 \times 10^{-2}$).

**PACS: 73.63.Fg**

The direct contact between carbon nanotube (CNT) and dielectric layer in CNT field effect transistors (CNT-FETs) affects both the transport[1,2] and noise characteristics[3]. Thus suspended CNT-FETs, in which direct contact with the dielectric is absent, have the potential for improved control and performance. However, fabrication methods for suspending CNTs demonstrated to date are arduous, and either involve harsh chemical treatment, or severely limit the materials which may be used. One approach is to selectively etch the dielectric layer underneath the CNTs to produce a suspended device[3]. This technique requires critical-point drying, and exposes the CNTs to harsh chemical processing. Another approach uses chemical vapor deposition (CVD) to grow CNTs on pre-patterned electrodes with trenches between the electrodes[1,2,4]. This method limits the substrate and electrodes to materials compatible with high temperature growth conditions. Here we present a simple, fast and scalable method of creating high-quality suspended CNT-FETs using a printing/lamination process. The technique does not expose the CNTs to any chemical treatment, and could be generalized to a wide variety of substrates and electrodes. The resulting FETs have improved transport and noise characteristics, and allow greater control of contacts to the source and drain.

The fabrication scheme involves transfer printing nanotubes[5,6] from a substrate covered with CVD-grown nanotubes to a second substrate with pre-patterned metal electrodes, as shown in Fig. 1(a). The source-drain electrodes (50 or 90 nm Au on 10 nm Ti, rms roughness 0.87 nm) are fabricated by a first step of photolithography on 300 nm thermal oxide with a p$^{++}$ Si substrate as a global back gate. Then a second step of photolithography is carried out to fabricate pillars (500 nm Au on 10 nm Ti, see schematic in Fig. 1(b)) on the inner edges of the electrodes. On a separate $SiO_2$ substrate, a CNT thin film (~ 150 CNTs/100 µm$^2$) is grown by a CVD process using an aqueous ferritin solution (0.5 mg/ml) as the catalyst[7]. The resulting nanotubes are mostly single-walled with average diameter around 1.4 nm. The nanotube covered substrate is pressed against the substrate with the electrodes at 300 psi and 150 °C for 3 minutes (Fig. 1(a)) using a nano-imprint machine (Nanonex NX2000). A SEM micrograph (Fig. 1(c)) of the electrodes after printing shows bright contrast for a suspended CNT, as confirmed by SEM imaging of a tilted sample. A CNT lying on SiO$_2$ across the channel and two short CNTs that did not bridge across the channel show darker contrast due to charge transfer[8]. A SEM image of the original CNT-covered SiO$_2$ substrate following printing (Fig 1(d)) shows 100% removal of CNTs in the region that was pressed into contact with the Au electrodes and approximately 4 % transfer of CNTs in the region of the channel between the electrodes. The pillars on the electrodes are essential to transfer printing. Their protrusion ensures conformal contact between the rough Au surface and the CNTs, and their size increases the local force between the substrates by reducing the area of contact. Increasing the height of the pillars causes decreased capacitance of the overall gate dielectric. Decreasing the height below 500 nm results in more CNTs touching the SiO$_2$; for example, with shorter 250 nm thick Au/Ti pillars most of CNTs fall on the SiO$_2$ surface. This configuration was used to make control FETs with a single CNT lying on SiO$_2$ in the channel. Decreasing the thickness of pillars further results in reduced transfer of CNTs to the pillars.

The number of suspended nanotubes in each device depends on channel length (*L*) and the original density of nanotubes on the growth substrate. The yield of devices with a single suspended CNT across the channel was optimized to 20% by making the channel width and channel length 5 µm and 1 µm, respectively. Three different growth densities (~ 35 CNT/100 µm$^2$, ~ 150 CNT/100 µm$^2$ and ~ 300 CNT/100 µm$^2$) of as-grown CNT thin films were tested, with 150 CNT/100 µm$^2$ resulting in the best yield of single-CNT devices. For these parameters, 60% of devices had multiple CNTs across the channel, whereas 20% of the devices had no CNTs. Both metallic and semiconducting nanotubes are observed following printing, and those with good semiconducting properties are used in the device characterization.

The number of suspended CNTs per device can be increased by repeating the transfer printing step multiple times. Figs. 2(a) and 2(b) show a device where the number of suspended CNTs was increased from 3 to 16 after 5 printing steps. The nanotube-covered SiO$_2$ substrate can be used several times for printing because the actual transfer of the CNTs occurs in the small area of the pillars, compared to the total area of the chip. Using this procedure, device characteristics can be tested for multiple configurations on the same electrode set. Furthermore, the printing process is also scalable to produce a large number of devices, as illustrated in Fig. 2(c). The SEM image shows a portion of a chip containing 51 electrode sets, onto which CNTs were simultaneously printed. Of the 51 sets, 10 devices were observed to have single suspended CNT, consistent with the expected 20% yield. In this case, a printing step takes 8 minutes to produce on average 10 suspended single-CNT devices, and the process could be further scaled up by printing on multiple chips simultaneously. Since our method is based on pressure-induced attachment of nanotubes to metal electrodes, suspended CNT-FETs can potentially be made on any robust substrate with a sufficiently malleable metal as the pillar-electrodes.

We have fabricated suspended CNT-FETs on SiO$_2$ and sapphire using Au and Pd as electrodes.

The quality of the printed suspended CNT-FET and control FET (with a single nanotube in contact with the SiO$_2$, see above) was tested using transport measurements at room temperature (Desert Cryogenics TT-Prober System) in air as well as in vacuum (5 x 10$^{-6}$ Torr). Suspended CNT devices have higher sub-threshold swing (> 1.3 V/decade), lower on-state conduction (< 1 μS), greater ambipolarity, smaller threshold voltage (gate voltage of minimum conductance), and less hysteresis than control CNT-FETs on SiO$_2$ (which are similar to those reported in literature[9]), as shown in Fig. 3(a). The differences likely result from a lack of doping from the substrate and/or reduced gate capacitance in suspended devices compared to control devices, as discussed below.

It has been noted previously that as-fabricated CNT-FETs on SiO$_2$ are p-doped, while suspended CNTs are nearly intrinsic[10]. Doping via the dielectric layer may also be removed by high-current annealing[11]. In this latter case, undoped CNT-FETs showed greater ambipolarity, smaller threshold voltage, and reduced on-state conductance, similar to the results presented here and consistent with our suspended CNT-FETs being undoped. Hysteresis is also reduced in the suspended CNT-FET as compared to the control device, consistent with hysteresis being due to charge traps on or in the dielectric[12-14].

The electrostatics of any suspended CNT device present some intrinsic issues with gate-coupling. Simulations of electric potential done using Poisson Superfish[15], as shown in Fig. 3(b), show that the contacts of both control (black line, 2) and suspended CNT-FETs (red line, 1) are only weakly coupled to the gate modulation, whereas the contacts in a standard top-electrode CNT-FET (blue line, 3) are strongly coupled. The simulations also confirm that suspended CNTs are more weakly coupled to the gate than control CNTs and standard top-electrode CNT-FETs, where the CNTs are physically closer to the gate. The decreased capacitance of the suspended CNT devices results in higher sub-threshold swing (> 1.3V/decade) compared to that reported in literature for CNT-FETs on SiO$_2$[9,11]. The fact that Schottky barriers at the contacts are relatively less influenced by gate modulation in case of both control and suspended CNT-FETs[16] may also play a role in the reduced on-conductance.

Another measure of device quality is the level of 1/$f$ noise in the CNT-FETs. Hooge's law expresses the low-frequency noise as $S_I = AI^\beta / f^\alpha$ [17], where $\beta = 2$, $\alpha \approx 1$, and $A = \alpha_H/N$ where $\alpha_H$ is Hooge's constant and $N = (c_g L |V_g - V_{th}|)/e$ is the number of carriers in the device. Previous investigations[18-24] of devices with the standard CNT-FET configuration (e.g. CNTs in contact with the oxide dielectric layer) have shown noise levels with an effective Hooge's constant of respectively (references[21], [20], [18]) $\alpha_H$ = 2 x 10$^{-3}$ and 9.3 x 10$^{-3}$ for diffusive transport and $\alpha_H$ = 7.4 x 10$^{-4}$ for ballistic transport. Here the current noise spectral density ($S_I$) was measured for suspended CNT-FETs and control devices under ambient conditions as well in vacuum (5 x 10$^{-6}$ Torr). Measurements were done in the linear region ($V_d << |V_g - V_{th}|$), at fixed source-drain bias, as a function of gate voltage.

Fig. 4(a) shows the frequency dependence of $S_I/I^2$ for two suspended devices and one control device at $V_d$ = 50 mV and $V_g$ = -30 V. The dependence of $S_I$ on frequency and current follows the expected Hooge's law behavior with $\alpha$ = 1 ± 0.14 and $\beta$ = 2 ± 0.12 for both the suspended and control CNT-FETs. The dependence of the noise on carrier density was evaluated from the slope of the inverse noise power, $I^2/S_I$, as a function of frequency ($I^2/S_I = (1/A)*f$) in the linear region of transport behavior of the device. The linear dependence of $1/A$ on $V_g$ shown in Fig. 4(b) indicates that mobility fluctuations rather than charge carrier number fluctuations are the origin for the noise in our suspended CNTs[20,25].

The noise power $S_I/I^2$ for the suspended CNT-FETs is 3 to 10 times smaller than that of the control device; this is especially surprising since the reduced gate capacitance means *smaller N* for the suspended device, which would give *larger* noise by Hooge's law. The noise power (3 x $10^{-8}$ to 8 x $10^{-9}$ at 100 Hz) is also smaller than that previously reported for suspended CNTs prepared by post-etching (~$10^{-7}$ at 100 Hz)[3]. To quantify the comparisons, the capacitance per unit length, $c_g$ for a long suspended CNT is calculated by $c_g \approx (2\pi k \varepsilon_o)/[\ln((h+2t)/h) + k\ln((2h^2+2ht-rh)/(rh+2rt))] =$ 8.2 aF/μm for the SiO$_2$ dielectric constant $k = 3.9$ and thickness $t = 300$ nm, the height of the CNT above the SiO$_2$ $h = 500$ nm and the CNT radius $r = 1$ nm. We estimate from numerical simulations that this capacitance is reduced to 2.6 aF/μm for $L = 1$ μm due to the screening by source-drain electrodes. Using this capacitance, the Hooge's constant for four suspended CNT devices ranged from 6.1 x $10^{-3}$ to 9.3 x $10^{-4}$ with an average of 2.6 x $10^{-3}$. In contrast, the capacitance of control device (geometry 2 in Fig. 3(b)) is calculated as 35.2 aF/μm, and the resulting Hooge's constant for 2 control devices was more than an order of magnitude larger, with an average of 5.6 x $10^{-2}$.

We have fabricated suspended nanotube devices with a straightforward technique that allows fabrication of as-grown nanotubes in a scalable fashion. The number of suspended CNTs can be controlled by tuning the parameters such as channel length and number of printing steps. Suspended CNTs showed reduction in substrate induced effects, demonstrating intrinsic ambipolar behavior with negligible hysteresis in vacuum. We find that suspending the nanotubes also decreases the spectral noise power by 3 to 10 times, and the average Hooge's constant for a suspended CNT-FET was found to be $2.6 \times 10^{-3}$, over an order of magnitude smaller than identical devices with the CNTs in contact with $SiO_2$. Such pristine suspended CNTs, with decreased influence of the $SiO_2$ surface both in transport characteristics as well as in $1/f$ noise, may enable studies of the intrinsic properties of carbon nanotubes.

**Acknowledgments:** This work has been supported by the Laboratory for Physical Sciences, and by use of the UMD-MRSEC Shared Equipment Facilities under grant # DMR 05-20471. Infrastructure support is also provided by the UMD NanoCenter and CNAM .

**Figure captions:**

Figure1. a) CNT covered substrate is pressed against electrodes at 300 psi, 90 °C for 3 minutes to transfer CNT in suspended configuration. b) Optimized dimensions of electrodes; the width of pillars (not shown) is 5 μm. c) SEM image of a device containing one suspended nanotube (indicated by arrow 1), one nanotube lying on $SiO_2$ (arrow 2) and two short CNT falling into the channel (arrow 3). d) SEM image of CNT covered substrate showing residual nanotubes after printing. Dashed line indicates the contact positions of the Au electrodes during printing.

Figure 2. a) 3 suspended nanotubes after one step of printing. b) 16 suspended nanotubes are obtained after five printing steps. c) SEM image of 5 electrode sets out of an array of 51 devices.

Figure 3. a) Transfer characteristics of a suspended CNT-FET (1) is compared with transfer characteristics of a control CNT-FET (2) in vacuum. Conduction is normalized with respect to on-state conduction, $G_{on}$: $G_{on,1}$ = 0.21 μS, $G_{on,2}$ = 2.2 μS. b) Simulations of equi-potential lines using Poisson-Superfish[15]. Three possible positions of CNTs are suspended (1), control (2) and top-electrode (3).

Figure 4. Noise measurements carried out in the linear regime, ($V_d$ << |$V_g$ −$V_{th}$|, where $V_d$, $V_g$ and $V_{th}$ are drain voltage, gate voltage and threshold voltage, respectively) using a current preamplifier (Ithaco 1211) and a spectrum analyzer (Stanford Research Systems SR 760). a) Inverse noise power as a function of frequency for a control device

and two extreme cases of suspended CNT-FETs at $V_d$ = 50mV and Vg = -30 V, i.e. when devices are fully turned on. b) 1/A as a function of $|V_g-V_{th}|$ for a suspended CNT-FET (# 2 from Fig 4(a)).

Figures:

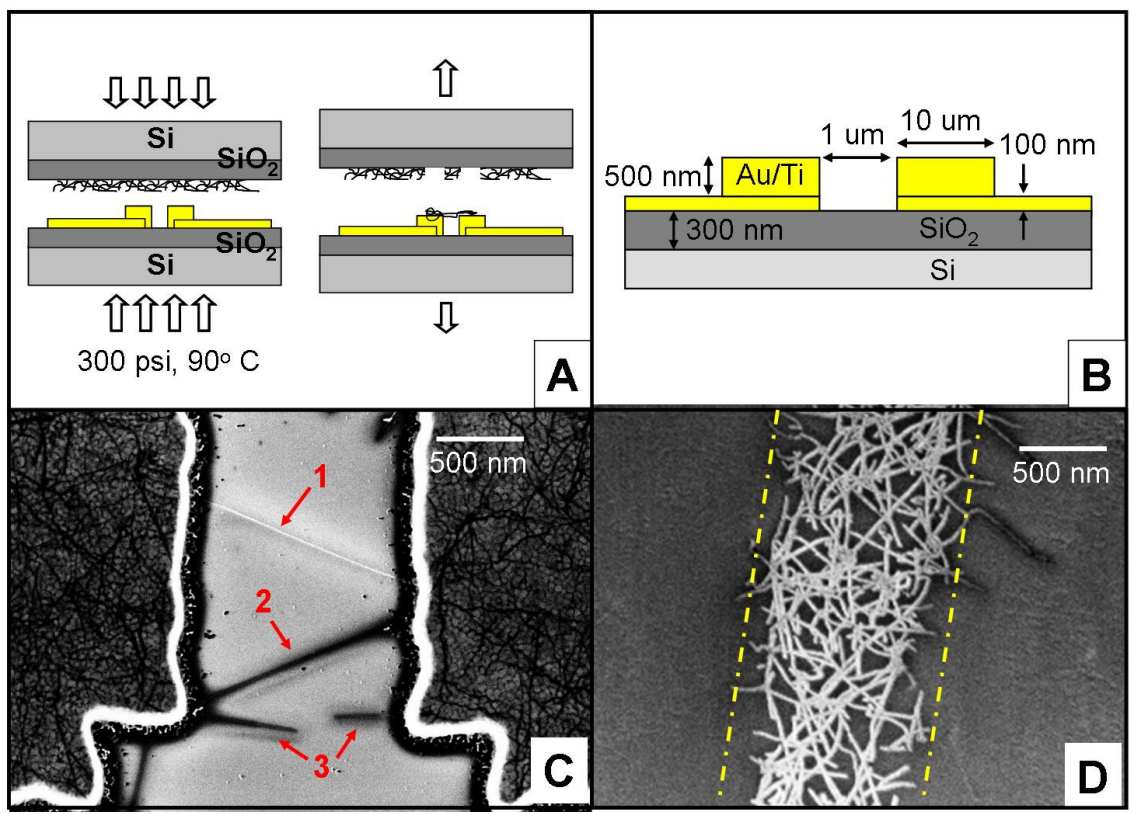

Figure1. A) CNT covered substrate is pressed against electrodes at 300 psi, 90 ºC for 3 minutes to transfer CNT in suspended configuration. B) Optimized dimensions of electrodes; the width of pillars (not shown) is 5 μm. C) SEM image of a device containing one suspended nanotube (indicated by arrow 1), one nanotube lying on $SiO_2$ (arrow 2) and two short CNT falling into the channel (arrow 3). D) SEM image of CNT covered substrate showing residual nanotubes after printing. Dashed line indicates the contact positions of the Au electrodes during printing.

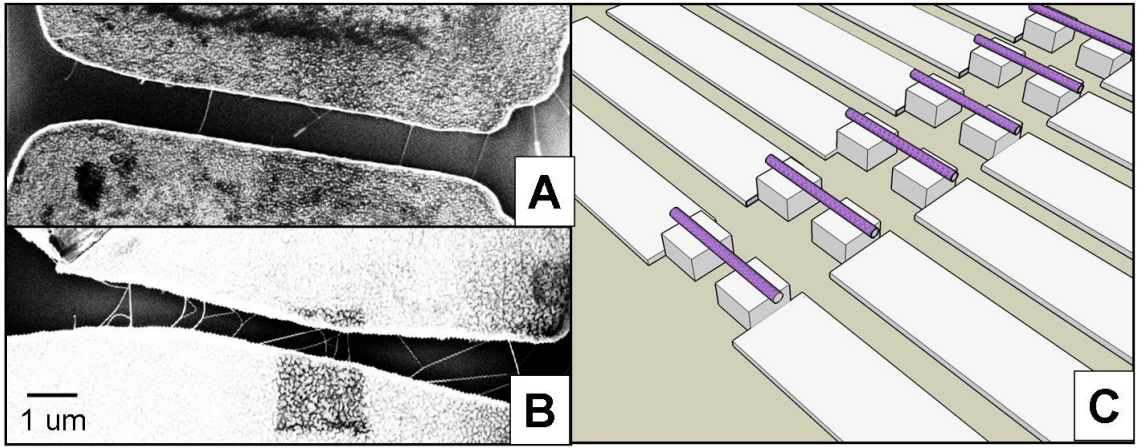

Figure 2. A) 3 suspended nanotubes after one step of printing. B) 16 suspended nanotubes are obtained after five printing steps. C) Schematic of array of pillared electrodes to scale up the fabrication process.

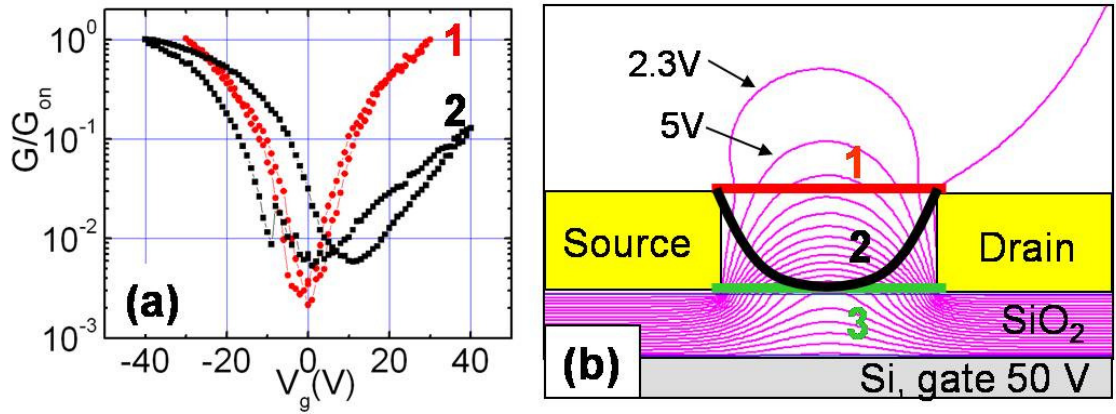

Figure 3. a) Transfer characteristics of a suspended CNT-FET (1) is compared with transfer characteristics of a control CNT-FET (2) in vacuum. Conduction is normalized with respect to on-state conduction, $G_{on}$: $G_{on,1}$ = 0.21 μS, $G_{on,2}$ = 2.2 μS. b) Simulations of equi-potential lines using Poisson-Superfish[15]. Three possible positions of CNTs are top-electrode (I), control (II) and suspended (III).

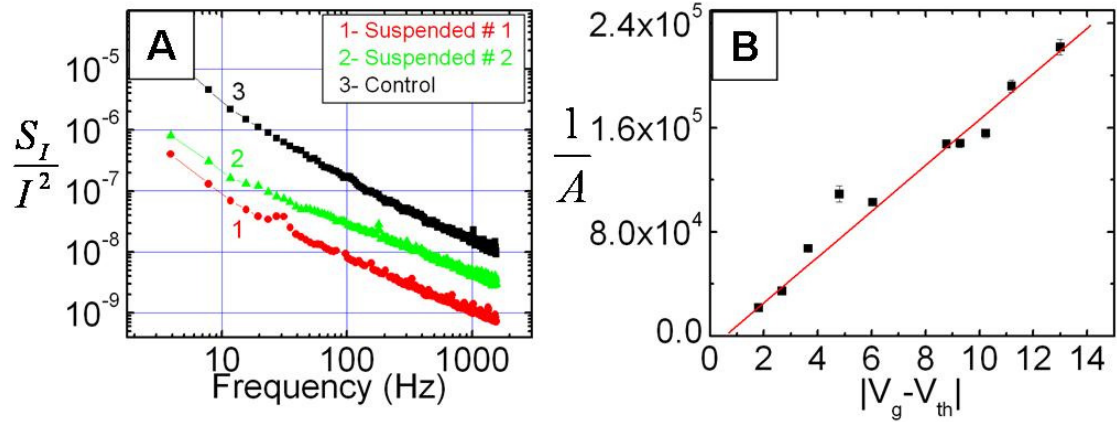

Figure 4. Noise measurements carried out in the linear regime, ($V_d \ll |V_g - V_{th}|$, where $V_d$, $V_g$ and $V_{th}$ are drain voltage, gate voltage and threshold voltage, respectively) using a current preamplifier (Ithaco 1211) and a spectrum analyzer (Stanford Research Systems SR 760). a) Inverse noise power as a function of frequency for a control device and two extreme cases of suspended CNT-FETs at $V_d$ = 50mV and Vg = -30 V, i.e. when devices are fully turned on. (b) 1/A as a function of $|V_g-V_{th}|$ for a suspended CNT-FET (# 2 from Fig 4(a)).